\documentstyle[twoside,fleqn,espcrc2,epsf]{article}
\newcommand\figcaption[1]{\vskip-0.38truein\caption{#1}\vskip0.1truein}

\newcommand\GeV{\mathord{\rm \;GeV}}

\newcommand\etal{{\it et al.}}

\begin{document}

\title{Lattice analysis of semi-leptonic form factors\thanks{Based on talks presented 
by Tanmoy Bhattacharya and Rajan Gupta.  These calculations have been done on
the CM5 at LANL as part of the DOE HPCC Grand Challenge program, and
at NCSA under a Metacenter allocation.}}

\author{Tanmoy Bhattacharya and Rajan Gupta
        \address{T-8 Group, MS B285, Los Alamos National
        Laboratory, Los Alamos, New Mexico 87545 U.~S.~A.~}
}

\begin{abstract}
We present preliminary results from simulations done on 170 $32^3
\times 64$ lattices at $\beta = 6.0$ using quenched Wilson fermions.
This talk focuses on the $Q^2$ behavior of the form-factors,
extrapolation in quark masses, dependence on renormalization scheme,
and comparison with heavy-quark effective theory (HQET). Even though
we cannot estimate errors due to quenching and discretization, our
results are consistent with experimental results for $D$ decays. We
present results for the Isgur-Wise function and estimate $\xi'(w=1) =
0.97(6)$.
\end{abstract}

\maketitle

\makeatletter 

\setlength{\leftmargini}{\parindent}
\def\@listi{\leftmargin\leftmargini
            \topsep 0\p@ plus2\p@ minus2\p@\parsep 0\p@ plus\p@ minus\p@
            \itemsep \parsep}
\long\def\@maketablecaption#1#2{#1. #2\par}

\advance \parskip by 0pt plus 1pt minus 0pt

\makeatother

\section{TECHNICAL DETAILS}
\label{s_q2dep}

We briefly mention some details of our analysis to extract the
form-factors at $Q^2=0$ relevant to phenomenology.  A full analysis
will be presented elsewhere \cite{MFFfinal}. The details of the
lattices are contained in the papers on the hadron spectrum
\cite{HM95} \cite{HMlat95}, and decay constants \cite{DC95}.  Preliminary results on
a sub-set of lattices have been presented at LATTICE 94
\cite{POTlat94} and DPF 94 \cite{dpf94}.  We do not have data to
extrapolate to $m_b$ or to $a=0$, thus our results are relevant for
$D$ decays, $i.e.$ $D\to Kl\nu$, $D\to\pi l \nu$, $D\to K^* l\nu$, and
$D \to \rho l \nu$, calculated at $a^{-1} = 2.33(4) \ \GeV$.

The decaying $D$ meson is created at rest by using a $\vec p = 0$
source.  On each lattice we make two measurements by creating the $D$
meson at $t=7$ and $57$, for a total statistical sample size of 340.
The sink for the final state meson is at $t=32$ in both cases.  The
time-slice of the weak operator is taken to vary between $10 \le t \le
30$ and $35 \le t \le 55$ respectively, and the insertion is at 5
lowest values of lattice momenta. The quark propagators are created
using a Wuppertal smeared source as described in \cite{HM95}.

\begin{figure}[t]
\figcaption{Ratio of 3-point to 2-point $SL$ function versus the time slice of the operator
insertion.}
\hbox{\epsfxsize=\hsize\epsfbox{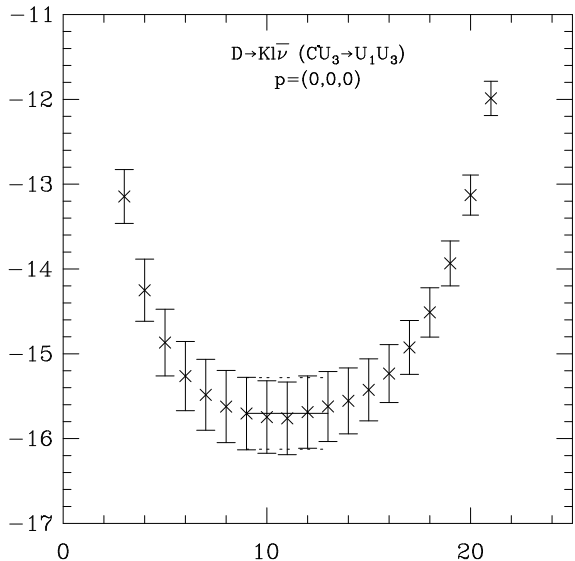}}
\vskip -18pt plus 10pt
\label{f_signal}
\end{figure}

In order to isolate the desired matrix element ($ME$) we construct a
ratio of 3-point to 2-point correlation functions. We have a choice of
using either smeared-smeared ($SS$) or smeared-local ($SL$) 2-point
correlation functions. We calculate the $ME$ both ways and take the
average as our best estimate. Figure~\ref{f_signal} shows a typical
example of signal for the ratio of correlators in the $SL$ case: the
quality in the $SS$ case is very similar. We have seen a steady
improvement in the consistency between these two estimates of $ME$ with
statistics. With the current sample they are within $1\sigma$ in all
cases.

\begin{figure}[t]
\figcaption{Three fits to test $Q^2$ behavior of $f_+$.}
\hbox{\epsfxsize=\hsize\epsfbox{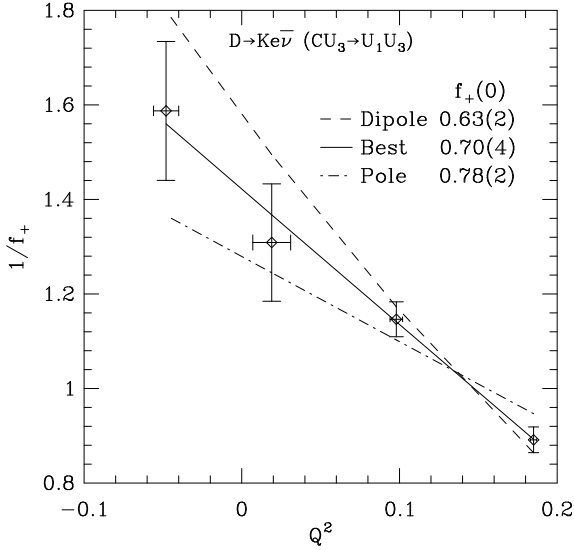}}
\vskip -24pt plus 10pt
\label{f_q2fits}
\end{figure}

\noindent{\bf Pole dominance hypothesis (PDH)}: It states that all
form-factors, $f(Q^2)$, have the structure
\begin{equation}
f(Q^2) \ = \ f(0) / (1 - Q^2/M^2)\,, \nonumber
\end{equation}
where $M$ is the mass of the nearest resonance with the right quantum
numbers.  To test PDH we make two kinds of fits: (i)
single parameter ``pole'' fit where $M$ is the lattice measured value
of the resonance mass, (ii) two parameter ``best'' fit where $M$ and
$f(0)$ are free parameters. An example of these fits is shown in
figure~\ref{f_q2fits}. Overall we find that only $f_0$ and $f_V$ are
well described by the ``pole'' form.  $f_+$ and $f_{A_0}$ are
consistent with ``pole'' form with $M < M_{pole}$, while for $f_{A_3}$
we find $M > M_{pole}$. $f_{A_1}$ and $f_{A_2}$ show a much smaller
$Q^2$ dependence than expected from pole dominance, however the data
are too noisy to make a definite statement.  We use results from ``best'' 
fit for our final estimates. 

\begin{figure}[t]
\figcaption{Comparison of $f_+(q^2)$ (squares) and $( 1 - {q^2 \over
(M_i + M_f)^2} )^{-1} f_0 (q^2)$ (diamonds) versus $w$.}
\hbox{\epsfxsize=\hsize\epsfbox{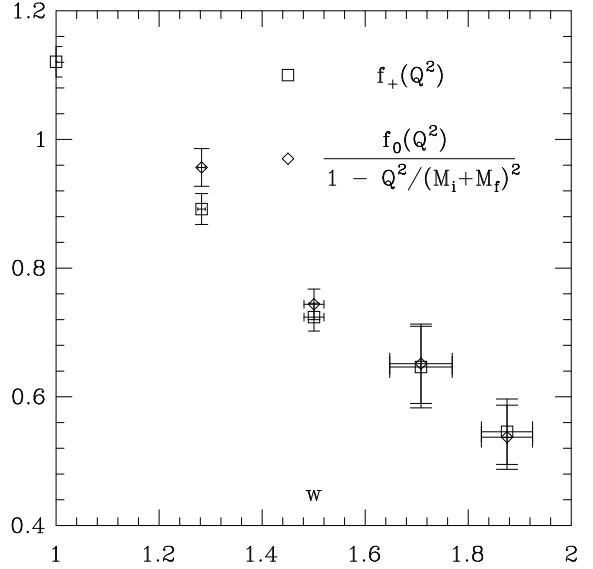}}
\vskip -18pt plus 10pt
\label{f_hqet}
\end{figure}

\noindent{\bf HQET}: At leading
order in $1/m_c$, HQET predicts that the Isgur-Wise function $\xi$
describes all form-factors relevant to $D \to K l \nu$ and $ D
\to K^* l \nu$ decays. Neglecting $O(\alpha_s)$ corrections, one gets \cite{neubert}
\begin{eqnarray}
\xi(w) &{} = R f_+(q^2) \ = R \left( 1 - {q^2 \over (M_i + M_f)^2} \right)^{-1} f_0 (q^2) 
             \aftergroup\hfill \nonumber \\
       &{} = R^* V(q^2)   \ = R^* A_0(q^2) \ = R^* A_2(q^2) \aftergroup\hfill \nonumber \\
       &{} = R^* \left( 1 - {q^2 \over (M_i + M_f)^2} \right)^{-1} A_1 (q^2) \,, \aftergroup\hfill\nonumber 
\label{e_hqet0}
\end{eqnarray}
where $M_i$ and $M_f$ are the initial and final meson masses and 
\begin{eqnarray}
  w  = v_i \cdot v_f = { M_i^2 + M_f^2 - q^2 \over 2 M_i M_f }; \ R 
     = {2 \sqrt{M_i M_f} \over M_i + M_f } \,.\nonumber 
\end{eqnarray}
The first HQET relation, $( 1 - q^2 / (M_i + M_f)^2 )^{-1} f_0 = f_+$,
is satisfied by our data as exemplified in figure~\ref{f_hqet} for $C
U_3 \rightarrow U_1 U_3$ decay.  Thus, both $f_0$ and $f_+$ cannot
simultaneously satisfy the PDH. If leading order HQET
holds, $i.e.$ $M_i+M_f \sim m_c + m_s \sim M_{pole}$, $f_+$ should
obey a `dipole' form as $f_0$ agrees with pole dominance.  The
data, as shown in figure~\ref{f_q2fits}, suggests that $f_+$ lies in
between ``pole'' and ``dipole'' forms.  Similar analysis for the
vector form factors is under progress.\looseness-1

\begin{figure}[t]
\figcaption{Extrapolation of $f_+(Q^2=0)$ to $m_u$. $f_+(D \rightarrow Kl\nu)$ is 
extracted from points labeled by squares and octagons, while $f_+(D \rightarrow \pi l\nu)$
is from data with degenerate $q \bar q$ points (crosses).}
\hbox{\epsfxsize=\hsize\epsfbox{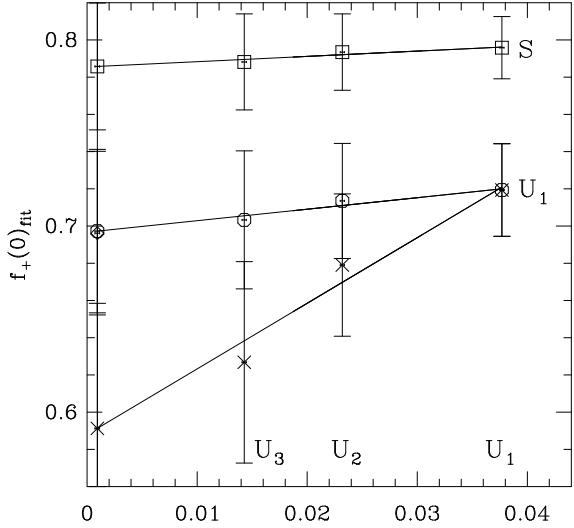}}
\vskip -18pt plus 10pt
\label{f_fpchiral}
\end{figure}

\noindent{\bf Dependence on quark mass}: 
Figure~\ref{f_fpchiral} shows an example of the variation of $f_+(0)$
with quark masses. There is significant dependence on the mass of the
quark $C$ decays into.  This is a kinematic effect as shown in
Section~\ref{s_iw}. Our data is good enough to expose slight
dependence on $m_{spectator}${}---the small decrease in slope between
the transitions $CU_i \to U_1U_i$ and $CU_i \to SU_i$ is consistent
with HQET.

\noindent{\bf Fixing \protect\boldmath $ m_s$}: We fix $m_s$ using $M^2_K / M^2_\rho$ 
and $M_\phi / M_\rho$. Our preferred way is $m_s(M_\phi)$, and the 
variation of results with $m_s$ is shown in Table~\ref{t_fq20}.

\noindent{\bf \protect\boldmath $O(ma)$ effects}: 
In Ref.~\cite{HM95} we discuss $O(ma)$ effects for heavy-light mesons.
The kinetic mass $M_2 \equiv d^2 E / d p^2$ is given by $\sinh M$ as
the data agree with the dispersion relation $\sinh^2{E/2} =
\sin^2{p/2} + \sinh^2{M/2}$.  We show variation of form-factors with
$M$ in Table~\ref{t_fq20}.

\begin{table}
\caption{Estimates of form factors in 3 commonly used 
renormalization schemes defined in \protect\cite{DC95}.}
\newcommand\ce[1]{\multicolumn{#1}{|c|}}
\begin{tabular}{|l|l|l|l|}
\hline
&TAD1&TAD$\pi$&TADU$_0$\cr\hline
$f_+        $&0.71(4) &0.75(4)  &0.66(4)  \cr
$f_0        $&0.73(3) &0.77(3)  &0.68(2)  \cr
$f_V        $&1.28(7) &1.36(7)  &1.19(6)  \cr
$f_{A_0}    $&0.84(3) &0.85(3)  &0.79(3)  \cr
$f_{A_1}    $&0.72(3) &0.74(3)  &0.68(3)  \cr
$f_{A_2}    $&0.49(9) &0.50(9)  &0.46(8)  \cr
$f_{A_3}    $&0.85(3) &0.87(3)  &0.80(3)  \cr
$f_V/f_{A_1}$&1.78(7) &1.84(8)  &1.76(7)  \cr
\hline
\end{tabular}

\vskip -24pt plus 10pt
\label{t_fqvsschemes}
\end{table}

\begin{table}
\caption{Estimates in $TAD1$ scheme at $Q^2=0$. The variations give 
estimates of systematic errors.}
\newcommand\ce[1]{\multicolumn{#1}{|c|}}
\setlength{\tabcolsep}{2.6pt}
\begin{tabular}{|l|l|l|l|l|l|l|}
\hline
&&\ce2{Pole}&\ce2{Best}\cr
\hline
&&$m_s(M_K)$&$m_s(M_\phi)$&$m_s(M_K)$&$m_s(M_\phi)$\cr
\hline
$f_+        $&$M_1$         &$0.78(2) $&$0.80(2) $&$0.70(4) $&$0.71(4) $\cr
$           $&$M_2$         &$0.79(2) $&$0.80(2) $&$0.67(5) $&$0.68(4) $\cr
\hline
$f_0        $&$M_1$        &$0.70(2) $&$0.72(2) $&$0.71(3) $&$0.73(3) $\cr
$           $&$M_2$        &$0.72(2) $&$0.73(2) $&$0.68(3) $&$0.70(2) $\cr
\hline
$f_V        $&$M_1$             &$1.27(7) $&$1.28(6) $&$1.27(7) $&$1.28(7) $\cr
$           $&$M_2$             &$1.22(6) $&$1.23(6) $&$1.22(9) $&$1.23(9) $\cr
\hline
$f_{A_0}    $&$M_1$             &$0.85(3) $&$0.85(3) $&$0.83(3) $&$0.84(3) $\cr
$           $&$M_2$             &$0.84(3) $&$0.84(3) $&$0.80(3) $&$0.81(3) $\cr
\hline
$f_{A_1}    $&$M_1$             &$0.67(2) $&$0.68(2) $&$0.71(3) $&$0.72(3) $\cr
$           $&$M_2$             &$0.65(2) $&$0.66(2) $&$0.67(4) $&$0.69(4) $\cr
\hline
$f_{A_2}    $&$M_1$             &$0.46(9) $&$0.47(8) $&$0.49(9) $&$0.49(9) $\cr
$           $&$M_2$             &$0.42(9) $&$0.44(9) $&$0.50(14)$&$0.52(12)$\cr
\hline
$f_{A_3}    $&$M_1$             &$0.84(3) $&$0.85(3) $&$0.84(3) $&$0.85(3) $\cr
$           $&$M_2$             &$0.81(3) $&$0.82(3) $&$0.81(4) $&$0.82(4) $\cr
\hline
$f_0\over f_+    $&$M_1$         &$0.93(1) $&$0.94(1) $&$1.02(4) $&$1.02(3) $\cr
$           $&$M_2$         &$0.95(1) $&$0.95(1) $&$1.03(4) $&$1.02(4) $\cr
\hline
$f_V\over f_{A_1}$&$M_1$             &$1.83(7) $&$1.81(6) $&$1.79(8) $&$1.78(7) $\cr
$           $&$M_2$             &$1.85(7) $&$1.83(6) $&$1.81(11)$&$1.79(10)$\cr
\hline
$f_{A_2}\over f_{A_1}$&$M_1$         &$0.70(10)$&$0.70(9) $&$0.69(12)$&$0.68(11)$\cr
$           $&$M_2$             &$0.65(11)$&$0.67(11)$&$0.74(18)$&$0.75(15)$\cr
\hline
\end{tabular}

\vskip -24pt plus 10pt
\label{t_fq20}
\end{table}

\noindent{\bf Renormalization Constants}: 
To relate lattice results to experimental data we need the 
renormalization constants $Z_A$ and $Z_V$.  We use three Lepage-Mackenzie 
tadpole improved schemes described in Ref.~\cite{DC95}. Our 
preferred scheme is $TAD1$, and the variation with the schemes is 
illustrated in Table~\ref{t_fqvsschemes}.

\section {RESULTS at $\beta=6.0$}

Our final results for $D \rightarrow Kl\nu$ and $D \rightarrow K^* l\nu$
are shown in tables~\ref{t_fqvsschemes}
and \ref{t_fq20} along with variation with $m_s$, type of fit, 
heavy-light meson mass, and the renormalization scheme.  Our preferred 
estimates are with $m_s(M_\phi)$, ``best'' fit, $M_1$, and $TAD1$ scheme. 
We also get that $f_+^\pi/f_+^K = 0.87(4)$. 

\section{\protect\boldmath $d \Gamma(Q^2)$ }

\begin{figure}[t]
\figcaption{$(1/V_{cs}^2) d \Gamma(q^2)/ d q^2$
versus $q^2$ in $\GeV^2$. The shape is in qualitative agreement with experimental data.}
\hbox{\epsfxsize=\hsize\epsfbox{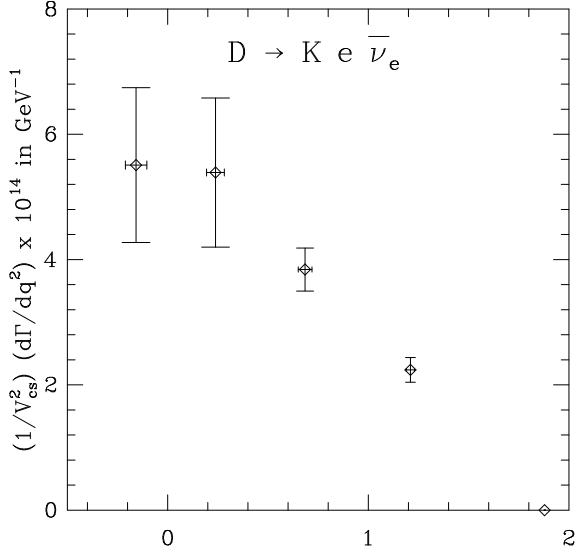}}
\vskip -18pt plus 10pt
\label{f_dgammaPS}
\end{figure}

\begin{figure}[t]
\figcaption{$(1/V_{cs}^2) d \Gamma(q^2)/ d q^2$
versus $q^2$ in $\GeV^2$.}
\hbox{\epsfxsize=\hsize\epsfbox{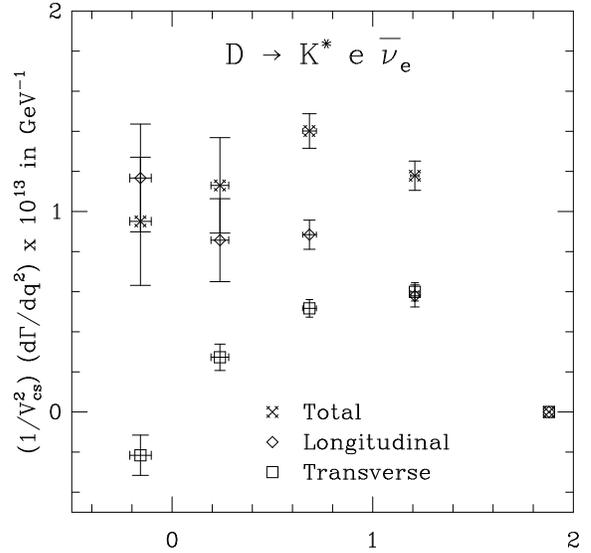}}
\vskip -18pt plus 10pt
\label{f_dgammaV}
\end{figure}

As explained in section~\ref{s_q2dep} there are considerable
uncertainties involved in extrapolating the form factors to $Q^2=0$.
Therefore, we also calculate $d\Gamma(Q^2) / d Q^2$ by linearly
extrapolating in $m_q$ the form-factors at fixed 3-momentum transfer.
In figure~\ref{f_dgammaPS} we show the results for $ D \rightarrow K e
\bar\nu_e$. In figure~\ref{f_dgammaV} we show the longitudinal,
transverse and the total decay widths for the process $ D \rightarrow
K^* e \bar\nu_e$.

\section{Isgur-Wise function}
\label{s_iw}

\begin{figure}[t]
\figcaption{$\xi(w)/\xi(1)$ at various quark masses and momenta. The symbols label the flavor
that the $C$ quark decays to, and variation of $\xi$ with $w$ is a
kinematic effect. Dependence on $m_{spectator}$ is shown most clearly
by the clusters of 3 points at $w \approx 1.2$. The largest value in
each cluster corresponds to the lightest spectator ($U_3$). The data are fit to
$\xi(w) = {2 \over w+1} \exp(-(2\rho^2-1){w-1 \over w+1})$, from which we estimate the 
slope $\xi'(1)$.}
\hbox{\epsfxsize=\hsize\epsfbox{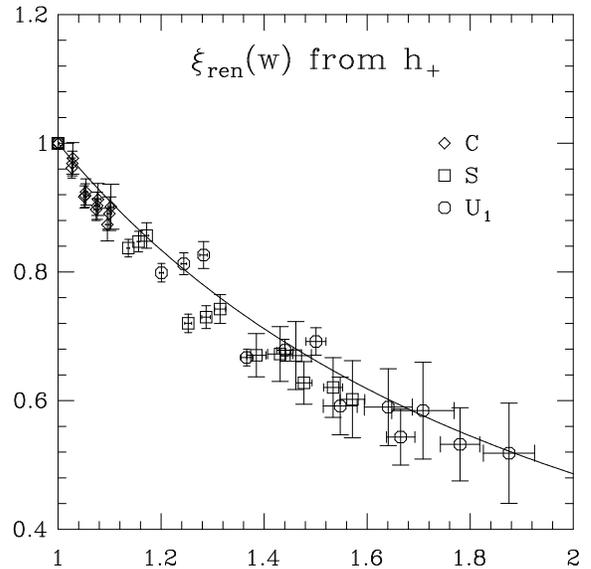}}
\vskip -12pt plus 10pt
\label{f_iw}
\end{figure}

As mentioned in section~\ref{s_q2dep}, at the leading order in the
heavy quark mass, there is one universal `Isgur-Wise' function $\xi$
which controls all the form factors. In figure~\ref{f_iw} we present
data for $\xi_{ren}(w)$,
\begin{eqnarray}
h_+(w) & {} = \left[ \hat C_{1} (w) +
                     {1 + w \over 2} ( \hat C_2 (w) + \hat C_3 (w) )
              \right] \xi_{ren} (w) \aftergroup\hfill \nonumber \\
       & {} = f_+(q^2)    { M_i + M_f \over 2 \sqrt{M_i M_f} }
                + f_-(q^2)    { M_i - M_f \over 2 \sqrt{M_i M_f} } ,\aftergroup\hfill \nonumber
\end{eqnarray}
where $\hat C_1$, $\hat C_2$, $\hat C_3$ are HQET renormalization constants 
including the leading $O(\alpha_s)$
corrections \cite{neubert}. The data show a slight dependence on the
spectator quark mass indicative of $m_c^{-1}$ corrections. Analysis 
from the vector form factors is in progress.  From this data we estimate 
the slope to be $\xi'(w=1) = 0.97(6)$. 

\end{document}